\documentclass{osa-article}
\journal{oe}
\usepackage{xcolor}
\begin{document}

\title{Localized surface electromagnetic waves in CrI$_3$-based magnetophotonic structures}

\author{Anastasiia A. Pervishko,\authormark{1} Dmitry Yudin,\authormark{1} Vijay Kumar Gudelli,\authormark{2} Anna Delin,\authormark{3,4,5} Olle Eriksson,\authormark{3,6} and Guang-Yu Guo\authormark{7,2}}

\address{\authormark{1}Skolkovo Institute of Science and Technology, Moscow 121205, Russia\\
\authormark{2}Physics Division, National Center for Theoretical Sciences, Hsinchu 30013, Taiwan\\
\authormark{3}Department of Physics and Astronomy, Materials Theory Division, Uppsala University, Box 516, SE-75120 Uppsala, Sweden\\
\authormark{4}Department of Applied Physics, School of Engineering Sciences, KTH Royal Institute of Technology, AlbaNova University Center, SE-10691 Stockholm, Sweden\\
\authormark{5}Swedish e-Science Research Center (SeRC), KTH Royal Institute of Technology, SE-10044 Stockholm, Sweden\\
\authormark{6}School of Science and Technology, \"{O}rebro University, SE-70182 \"{O}rebro, Sweden\\
\authormark{7}Department of Physics and Center for Theoretical Physics, National Taiwan University, Taipei 10617, Taiwan}

\begin{abstract}
Resulting from strong magnetic anisotropy two-dimensional ferromagnetism was recently shown to be stabilized in chromium triiodide, CrI$_3$, in the monolayer limit. While its properties remain largely unexplored, it provides a unique material-specific platform to unveil its electromagnetic properties associated with coupling of modes. Indeed, trigonal symmetry in the presence of out-of-plane magnetization results in a non-trivial structure of the conductivity tensor, including the off-diagonal terms. In this paper, we study the surface electromagnetic waves localized in a CrI$_3$-based structure using the results of {\it ab initio} calculations for the CrI$_3$ conductivity tensor. In particular, we provide an estimate for the critical angle corresponding to the surface plasmon polariton generation in the Kretschmann-Raether configuration by a detailed investigation of reflectance spectrum as well as the magnetic field distribution for different CrI$_3$ layer thicknesses. We also study the bilayer structure formed by two CrI$_3$ layers separated by a SiO$_2$ spacer and show that the surface plasmon resonance can be achieved at the interface between CrI$_3$ and air depending on the spacer thickness.
\end{abstract}

\section{Introduction}
Studies of two-dimensional materials have been booming since the successful exfoliation of the monolayer graphene structure \cite{Novoselov2004}. This pioneering work initiated the profound study of its electronic, transport and optical properties. In particular, it was shown that graphene with a Dirac-like spectrum can host the surface electromagnetic waves \cite{ju2011graphene,grigorenko2012graphene, garcia2014graphene} and also the edge plasmon waves \cite{PhysRevB.85.235444, Kumar2016} that have been proposed as key ingredients for different potential applications such as ultrafast optical modulators \cite{liu2011graphene}, biosensors \cite{wu2010highly, rodrigo2015mid}, and photodetectors \cite{liu2011plasmon, vicarelli2012graphene}. Along with that the search for alternative two-dimensional structures with hexagonal lattices and similar properties favoring their potential utility has been started. Lately, the experimental activity has resulted in the discovery of other members of this family, including monolayers of transition metal dichalcogeniedies \cite{Manzeli2017}, phosphorene \cite{Li2014,Liu2014}, silicene \cite{Vogt2012,Feng2012,Fleurence2012}, germanene \cite{Davila2014} and stanene \cite{Zhu2016} to name a few. Targeted data-mining efforts have also been devoted to the identification of novel two-dimensional materials \cite{Lebegue2013,Eriksson2018,Mounet2018}, and in fact some of these theoretical predictions have been verified experimentally \cite{Romdhane2015,Lin2016, Gong2017}.

Of particular importance is the recent discovery of intrinsic ferromagnetism in Cr$_2$Ge$_2$Te$_6$ and CrI$_3$-based van der Waals structures, where magnetism retains down to a monolayer limit as has been shown using magneto-optical Kerr effect \cite{Huang2017,Gong2017}. The observed phenomena have opened avenues for the direct applications of this class of materials, for ultrathin magnetic sensors and spin filters of high efficiency \cite{Zhong2017,Seyler2018,Jiang2018,Klein2018,Wang2018,Huang2018} and in turn motivated the present study. For a long time magnetism was not believed to be present in low dimensional magnetic structures, owing to the Mermin-Wagner theorem \cite{Mermin1966}, which in the isotropic Heisenberg model forbids the formation of collinear magnetic ordering in one- and two-dimensional systems, at any finite temperature. Nevertheless, as confirmed experimentally magnetic ordering in CrI$_3$ crystals occurs due to the large out-of-plane magnetic anisotropy that relaxes the Mermin-Wagner constraint. The magnetic anisotropy turns out to substantially suppress transverse spin fluctuations resulting in ferromagnetic ordering being clearly detectable at finite temperatures, even in monolayer systems. 

Experimental findings on CrI$_3$-based monolayers, including magnetic exchange coupling and magnetic anisotropy, are supported by the results of first-principles calculations \cite{Zhang2015,Lado2017,Torelli2018}. The magnetic properties of this system are governed by the Cr$^{3+}$ ions, which are arranged in a two-dimensional hexagonal Bravais lattice, while the non-magnetic I$^-$ ions form an octahedral coordination around Cr. The combination of both relativistic spin-orbit coupling of the heavier ligand I atoms, as well as single-ion anisotropy of Cr atoms, leads to the strong magnetic anisotropy of the structure \cite{Zhang2015,Lado2017,Jiang2018a,Fang2018,Andersson2007}. There is a huge surge of interest in novel two-dimensional materials. Indeed, characteristic energy scales for electron correlations and magnetism are the same in two dimensions, making thus the interplay between them especially strong in newly discovered structures. Engineered from a variety of magnetic layers and controlled by an external electromagnetic field or electric current, low-dimensional magnets are expected to yield a plethora of new magnetic orders with interesting physical properties. They provide the most promising playground for building novel and energy efficient memory chips as well as quantum information and neuromorphic devices. Unfortunately, from this perspective, not much is known regarding electromagnetic properties of chromium triiodide monolayers. To remedy some of this shortcoming, in this paper we explore mono- and nanolayers of chromium triiodide from the electromagnetic theory perspective with the focus on surface plasmon excitations. The formation and propagation of surface electromagnetic waves at the interface between two media has been extensively studied over last decades \cite{Pitarke2006,Schuller2010,Polo2011}. Initiated by the description of surface waves at the boundary between the metal and dielectric material, their presence has also been confirmed on specific semiconductor/dielectric interfaces, including a whole family of two-dimensional materials \cite{Mikhailov2007,Iorsh2013,Yudin2015,Kumar2016,Liu2016,Yudin2016,Yudin2017,Bludov2019}. In this paper, we investigate the possibility of surface electromagnetic wave generation based on the solution to a set of Maxwell's equations with material specific conductivity tensor as obtained from {\it {ab initio}} calculations. We address several systems containing CrI$_3$ layers, in order to explore the potential of surface plasmon generation in this class of materials and discuss the optimal conditions for their formation.

\begin{figure}
\centering
\begin{minipage}{.5\textwidth}
  \centering
  \includegraphics[width=\linewidth]{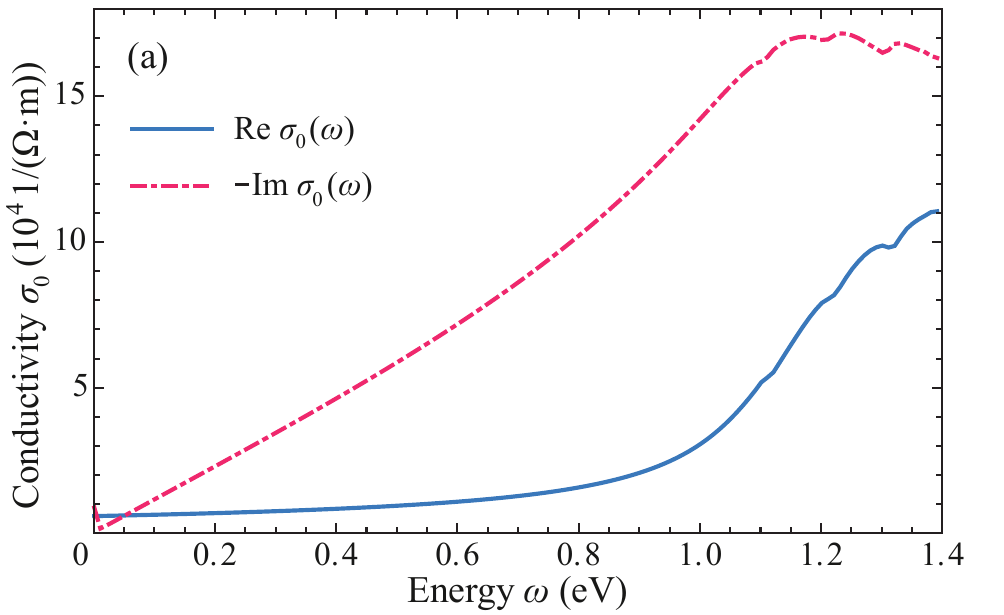}
\end{minipage}%
\begin{minipage}{.5\textwidth}
  \centering
  \includegraphics[width=\linewidth]{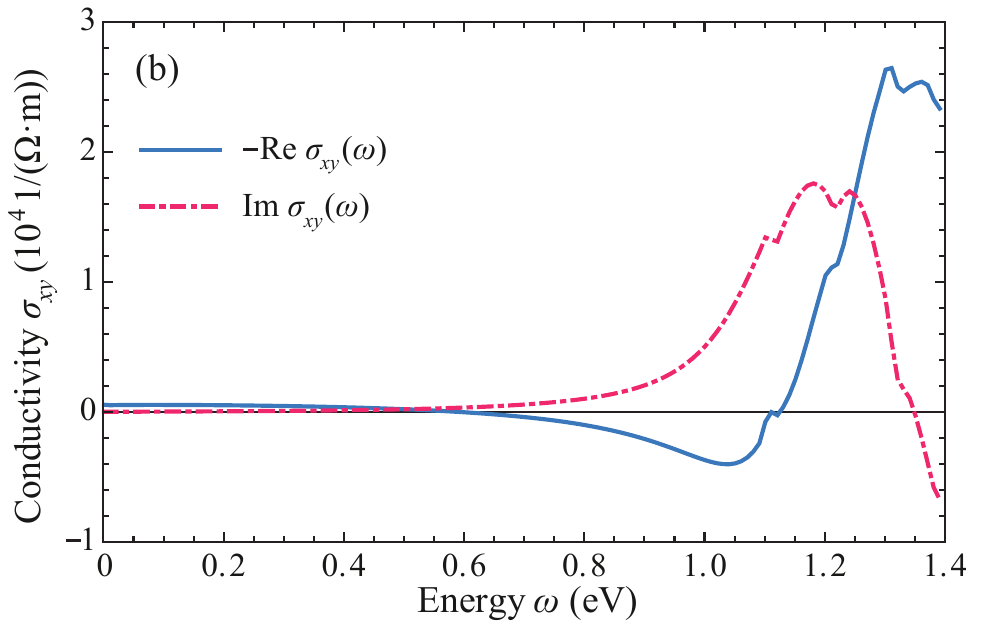}
\end{minipage}
\caption{In-plane (a) diagonal $\sigma_{0}(\omega)$ and (b) off-diagonal $\sigma_{xy}(\omega)$ components of the conductivity tensor as obtained from the Kubo formula combined with the results of first principles calculations within GGA+$U$ approximation \cite{Gudelli2019}.}
\label{fig:conductivity}
\end{figure}

\section{Optical conductivity tensor}

For the calculation of the conductivity tensor of a CrI$_3$-monolayer we applied density functional theory, with the use of the accurate, fully relativistic projector augmented wave (PAW) method, as implemented in the Vienna ab-initio simulation package (VASP) \cite{Kresse1996,Kresse1996a}. This allows us to trace back the effect of spin-orbit interaction as obtained from electronic band structure calculations. Indeed, spin-orbit interaction is known to be an effective mechanism that couples electronic and magnetic degrees of freedom. In the absence of an external magnetic field, the electric field affects magnetization via spin-orbit torques which are known to be specified by conduction electrons exclusively. Therefore, the effect of magnetization is taken into account by the first principles simulations and, in combination with spin-orbit coupling, it influences the values of the off-diagonal contribution of the conductivity tensor (see, e.g. \cite{Shibata2018}). Exchange-correlation potential in form of the Perdew-Burke-Ernzerhof parametrization \cite{Perdew1996} is applied within the generalized gradient approximation (GGA), with the account of on-site Coulomb repulsion among Cr 3$d$ electrons (GGA+$U$) \cite{Jeng2004,Dudarev1998}. Similar to the previous study on Cr$_2$Ge$_2$Te$_6$ \cite{Fang2018}, we adopt $U=1$ eV. From these calculations, a material specific conductivity tensor can be worked out based on the Kubo formula \cite{Adolph2001}. Without loss of generality, assume that CrI$_3$ monolayer is placed at $z=0$ plane. A proper account of trigonal symmetry of a honeycomb lattice suggests only three components to be independent, namely $\sigma_0(\omega)\equiv\sigma_{xx}(\omega)=\sigma_{yy}(\omega)$, $\sigma_{yx}(\omega)=-\sigma_{xy}(\omega)$, and $\sigma_{zz}(\omega)$ with the rest being zero. The results of the GGA+$U$ calculations \cite{Gudelli2019} for in-plane components of the conductivity tensor, $\sigma_0(\omega)$ and $\sigma_{xy}(\omega)$, are shown in Fig.~\ref{fig:conductivity}.

Detailed analysis of the CrI$_3$ band structure calculated using GGA+$U$ method (shown in Fig. 7 in Ref.~\cite{Gudelli2019}) allows quantitatively identify the origin of the absorptive components of the conductivity, $\mathrm{Re}\,\sigma_0(\omega)$ and $\mathrm{Im}\,\sigma_{xy}(\omega)$. As it was found, they are directly linked to the dipole allowed interband transitions in the system, which is in full agreement with the experimental findings \cite{Huang2017}. The peak in Fig.~\ref{fig:conductivity}b is associated with the transition from the occupied state with $-0.1$ eV to the unoccupied one with $1.1$ eV, while the peak at around $1.5$ eV in Fig.~\ref{fig:conductivity}a is attributed to interband transition from the valence band at $-0.4$ eV to the conduction band at $1.1$ eV. A detailed analysis of optical transitions within GGA+$U$ approach can be found in Ref.~\cite{Gudelli2019}. 

It should be emphasized that within GGA+$U$ approximation the conductivity of a CrI$_3$ monolayer does not differ qualitatively from the behavior of bulk, as reported in Ref.~\cite{Gudelli2019}, and therefore the results presented in Fig.~\ref{fig:conductivity} can be generalized to situations involving thicker CrI$_3$ layers. The conductivity tensor of CrI$_3$ can be also found  from calculations based on the $GW$-BSE approximation. This level of approximation allows to take into account electron-hole interactions, that are neglected in GGA+$U$ approximation. We also provide such analysis below, but we note here that qualitatively these results agree with each other.
  
A close inspection of Fig.~\ref{fig:conductivity} clearly reveals that for energies $\omega\lesssim1.2$ eV the off-diagonal component of the conductivity tensor $\sigma_{xy}(\omega)$  is considerably smaller than the diagonal one. In addition to that, it is interesting to note that for small energies, the imaginary part of the diagonal component of the conductivity tensor is negative while the real part is small. Therefore the surface plasmon waves can potentially be generated in CrI$_3$ layer, and this possibility will be discussed below.

\section{Results and discussion}

\subsection{Dispersion of surface electromagnetic waves} 
\begin{figure}
    \centering
    \includegraphics[scale=0.5]{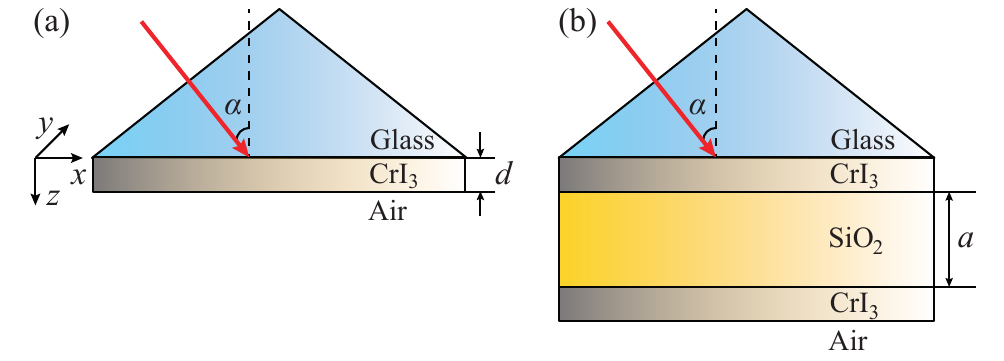}
    \caption{The system under consideration: (a) the CrI$_3$ layer of thickness $d$ is placed on top of a glass prism with permittivity $\varepsilon_1$, while the air permittivity is $\varepsilon_2$. The light is incident at the interface between glass and CrI$_3$ at the angle $\alpha$; (b) the bilayer structure is formed by two thin CrI$_3$ layers separated by a SiO$_2$ layer of thickness $a$, with refractive index $n_{\mathrm{SiO}_2}$. Light is injected into the structure via a glass prism at the angle $\alpha$.}
    \label{fig:model}
\end{figure}

Generation, propagation and detection of plasmon excitations, resulting from the coupling between electromagnetic field and charge density waves of a conducting media, are of central importance of the rapidly advancing area of plasmonics \cite{Schuller2010}. In this context, chromium triiodide with its unique magneto-optical properties serves as a candidate for hosting surface electromagnetic waves formed in this monolayer structure. To outline the mechanism underlying the emergence and stability of surface plasmon polaritons in a CrI$_3$ monolayer, we inspect Maxwell$’$s equations in the geometry shown schematically in Fig.~\ref{fig:model}a. The magnetic CrI$_3$ layer is positioned at $z=0$ with the $\hat{\it{z}}$ axis pointing out from an insulating medium represented by a glass prism with constant permittivity $\varepsilon_1$ ($z<0$). Outside the CrI$_3$ layer is air with permittivity $\varepsilon_2=1$.  Without loss of generality, we suppose the electromagnetic wave propagates along the $\hat{\it{x}}$ axis with the propagation constant $q$ and consider an evanescent solution which decays exponentially along the $\hat{\it{z}}$ axis, $\propto e^{-\kappa|z|+iqx}$.

\begin{figure}
    \centering
    \includegraphics[scale=0.5]{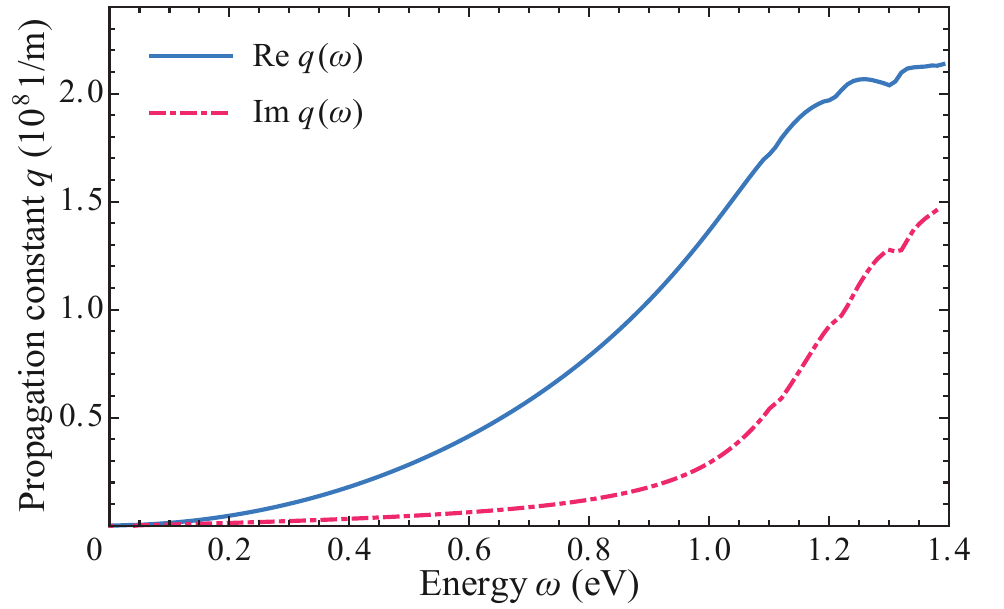}
    \caption{The dispersion relation of surface electromagnetic waves $q(\omega)$ obtained as the solution to Eq.~(\ref{disp}). It is clearly visible that in the energy window up to 1.2 eV the imaginary part of propagation constant is almost zero $\mathrm{Im}~q(\omega)\approx 0$, favoring the generation of surface plasmon polaritons along the CrI$_3$ boundary.}
    \label{fig:dispersion}
\end{figure}

Likewise other conducting materials, the presence of a chromium triiodide monolayer results in the discontinuity of the magnetic field at the boundary $z=0$, $\hat{\it{z}}\times(\it{H}_2-\it{H}_1)=\hat{\sigma}(\omega)\it{E}^\mathrm{in}$, while leaving the in-plane components of the electric field $\it{E}^\mathrm{in}=(E_x,\,E_y)$ unchanged, $\it{E}_1^\mathrm{in}=\it{E}_2^\mathrm{in}$. Thus, under these specified boundary conditions, one may obtain the expression for the dispersion \cite{Chiu1974,Iorsh2013,Yudin2015,Kumar2016}:
\begin{equation}\label{disp}
    \left(\frac{i\sigma_0}{\omega\varepsilon_0}+\frac{\varepsilon_1}{\kappa_1}+\frac{\varepsilon_2}{\kappa_2}\right)\left(\frac{i\sigma_0}{c\varepsilon_0}k_0-\kappa_1-\kappa_2\right)=\left(\frac{\sigma_{xy}}{c\varepsilon_0}\right)^2,
\end{equation}
where $k_0=\omega/c$, while $\varepsilon_0$ and $c$ denote the vacuum permittivity and speed of light respectively, and with  $\kappa_{1,2}^2=q^2-\varepsilon_{1,2}k_0^2$ on either side of the monolayer. In Eq.~(\ref{disp}) multipliers on the left-hand side stand for the dispersion of a transverse electric (TE) and a transverse magnetic (TM) modes that can potentially be achieved in the selected media. Note that when the off-diagonal component $\sigma_{xy}=0$ the expression ~(\ref{disp}) decouples and represents the superposition of two independently propagating waves. However, due to the presence of the off-diagonal component for CrI$_3$ monolayer, the right-hand side of Eq.~(\ref{disp}) results in the electromagnetic wave being of hybrid nature due to the mixing between TE and TM modes proportional to $\sigma_{xy}(\omega)$ squared.

Numerical solution to Eq.~(\ref{disp}) allows to identify the dispersion of localized surface electromagnetic waves, as shown in Fig.~\ref{fig:dispersion}. The imaginary part,  $\mathrm{Im}~q(\omega)$, defines the attenuation constant and is almost negligible in a relatively wide spectral region (up to $\omega\simeq1.2$ eV), thus favoring the formation and propagation of surface plasmon polaritons bound to the ferromagnetic CrI$_3$ monolayer. However, the imaginary part of $\mathrm{Im}~q(\omega)$ becomes rather pronounced for higher frequencies, which takes place in the vicinity of interband transitions corresponding to the absorption edge. Remarkably, upon inverting the plot in Fig.~\ref{fig:dispersion} we reproduce the well-known square-root behavior at small momenta, $\omega\propto\sqrt{q}$, with proportionality parameter 21.883 GHz $\cdot$ m$^{-1/2}$.

\subsection{Electromagnetic modeling using GGA+$U$ calculations} 
To prove the presence of surface electromagnetic waves bound by a magnetic nanolayer, we simulate the propagation of TM-polarized electromagnetic field through a thin film of chromium triiodide in the  Kretschmann-Raether (KR) configuration (Fig.~\ref{fig:model}a) \cite{Kretschmann1968,Zayats2005,Gray2013,Foley2015,Akimov2017,Vinogradov2018}. This method is known to be quite ubiquitous in generating surface plasmon polaritons. The standard setup consists of a glass prism and a thin film of a lossy material. In our study the latter is the chromium triiodide layer, of thickness $d$, covered by an insulating medium that has a lower refractive index, compared to the glass prism on the opposite side of the chromium triiodide layer. For incident angles greater than the angle of the total internal reflection, the light reaching the boundary between the prism and the thin film material is converted to the evanescent wave at the other side of the boundary, thus ensuring the coupling to the surface electromagnetic waves. The maximum coupling occurs when the wave vector of the incident light matches the value of the surface plasmon propagation constant. In the reflectance spectrum, the excitation of the surface plasmon polaritons manifests itself as a dip of the reflectance curve, taking place at higher angles compared to the peak representing the effect of total internal reflection.

To study light propagation through the system described in Fig.~\ref{fig:model}a, we solve numerically a set of Maxwell$'$s equations with account for electromagnetic boundary conditions at the surface and CrI$_3$ layer thickness. We consider a TM-polarized field, incident on the prism at the angle $\alpha$, that is characterized by the following parameters: $\left\vert H_y\right\vert=1$ A/m, the frequency is $f_0=242$ THz, and input power 1 W. This frequency has been chosen to make the effect as pronounced as possible in the simulations. The refractive indices on both sides of the chromium triiodide monolayer were taken to be: $n_1$=1.5 (for glass) and $n_2$=1 (for air). For this energy of the field, the components of the nanolayers conductivity tensor were calculated to be: $\sigma_0=(0.31-1.42i)\times10^5$ $\Omega^{-1}\cdot$ m$^{-1}$, $\sigma_{xy}= (0.37+0.51i)\times 10^4$  $\Omega^{-1}\cdot$ m$^{-1}$ (see Fig.~\ref{fig:conductivity}).
\begin{figure}
\centering
  \includegraphics[scale=0.5]{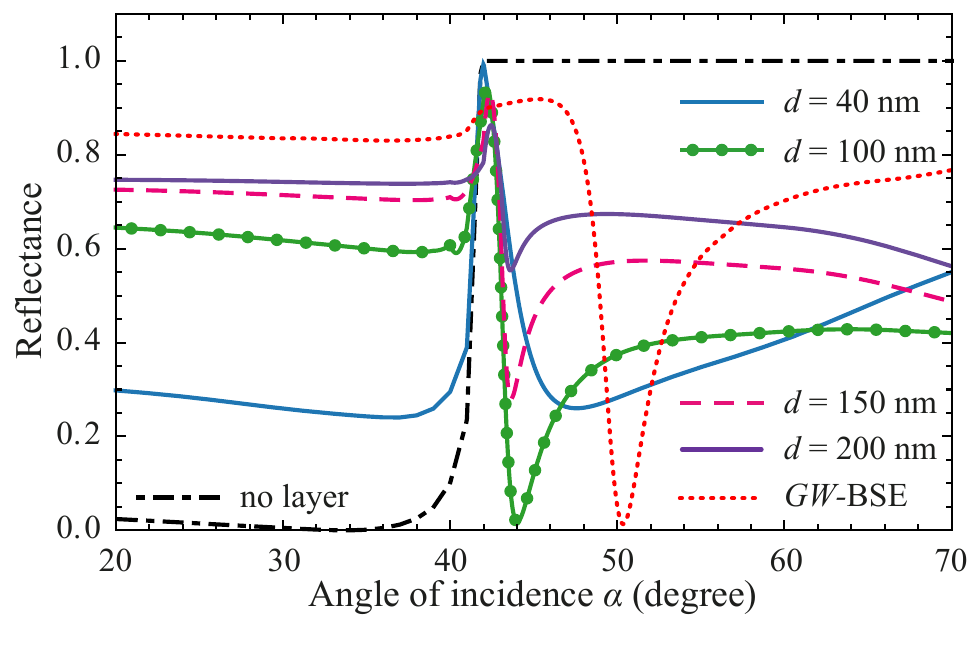}
  \caption{Reflectance spectrum, characterizing the ratio of the reflected power to the input power (we put 1 W throughout the simulations) in the glass/CrI$_3$/air Kretschmann-Raether configuration (see Fig.~\ref{fig:model}a) given by the exact solution to Maxwell's equations with $\lambda=1238$ nm for different CrI$_3$ layer thicknesses, $d$. For $d=100$ nm the curve drops down to almost zero at $\alpha_0=44.1^\circ$ (green solid-circled line), which corresponds to the surface plasmon resonance while the total internal reflection of the incident electromagnetic field in the glass/air system without CrI$_3$ layer takes place at $\alpha=41.8^\circ$ (black dash-dotted line). For $d>100$ nm the maximum reflectance peak decreases and becomes narrower with the increase of the layer thickness. It almost vanishes at $d=200$ nm (violet solid-squared line), while for $d<100$ nm the surface plasmon resonance is not profound since there is not enough thickness to absorb the incident light. The reflectance spectrum of CrI$_3$ nanolayer of the thickness $d$=300~nm for $\lambda=886$~nm within $GW$-BSE methodology is shown by red dotted line.}
  \label{fig:reflectance}
\end{figure}

The results of numerical simulation of the reflectance spectrum for different CrI$_3$ layer's thicknesses are shown in Fig.~\ref{fig:reflectance}, where the dips of the reflectance curve stand for the effect of total absorption of the electromagnetic field energy, which is accompanied by the surface plasmon generation. A close inspection of these results reveals that the resonance angle of surface plasmon polaritons for glass/CrI$_3$/air KR configuration ranges from $\alpha=43.6^\circ $ to $47.8^\circ$ for $d$ between 40 and 200 nm whereas the angle of total internal reflection for the system glass/air is $\alpha=41.8^\circ$. Comparing the position of the surface plasmon dip, one can see that the thickness of the CrI$_3$ layer, $d$, is an important parameter for surface plasmon generation in the investigated system. The optimal thickness, resulting in almost zero reflectance value ($\approx 0.005$ a.u.), equals $d=100$ nm. This demonstrates a very efficient excitation of surface plasmons at these conditions, while for the thinner CrI$_3$ layers the surface plasmon resonance is less pronounced because there is not enough thickness to absorb light and excite plasmons. Subsequent variation of the CrI$_3$ layer thickness ($d>100$ nm) results in a narrowing and a decrease of the resonance peak amplitude, which is due to optical losses.  

\begin{figure}
\centering
  \includegraphics[scale=0.5]{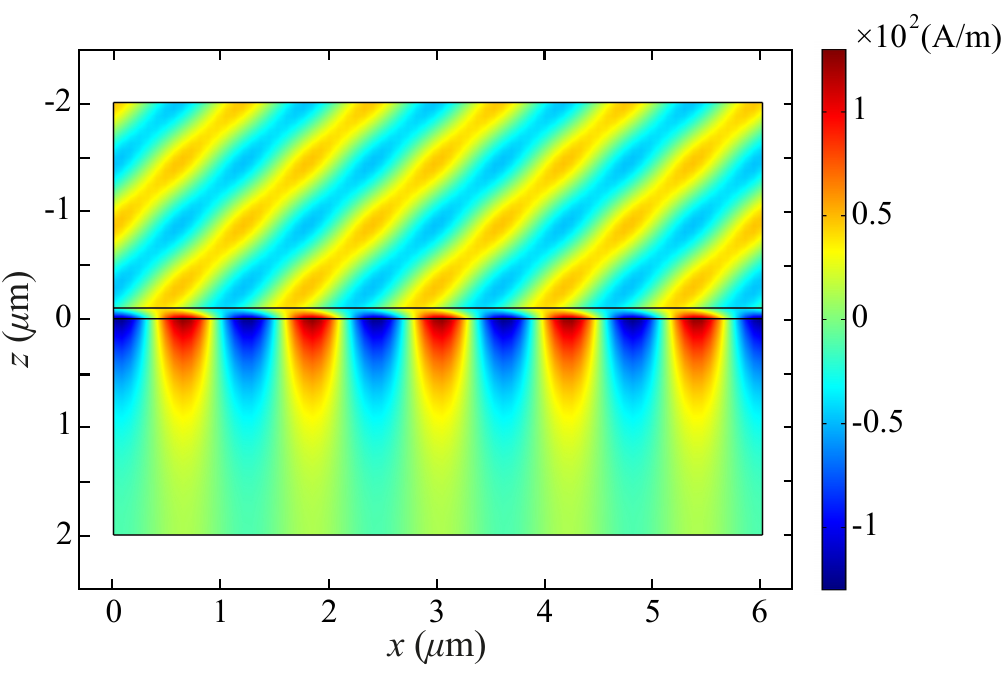}
  \caption{Spatial distribution of magnetic field component $H_y$ of TM polarized light through the system at the incident angle $\alpha=44.1^\circ$, showing the propagation of surface electromagnetic waves bound by a CrI$_3$ layer with thickness $d=100$ nm positioned at the $z=0$ plane and its exponential decay in the insulating medium.}
 \label{fig:2d_surf}
\end{figure}

  In Fig.~\ref{fig:2d_surf}, we present the color contour plot of the $y$-component of the magnetic field,  $H_y$, and its distribution in the {\it{xz}}-plane for a TM-polarized electromagnetic field that propagates across the system with optimal thickness, $d=100$ nm, at an incident angle $\alpha=44.1^{\circ}$. The profile of the magnetic field component is clearly revealing an electromagnetic field that is localized at the chromiun triiodide layer, with an evanescent nature in the insulating medium. 

\subsection{Electromagnetic modeling using $GW$-BSE calculations}\label{app:appa}

In the previous section, we analyzed in detail surface electromagnetic waves with the conductivity tensor as obtained from GGA$+U$ methodology \cite{Gudelli2019}. Such an approach is known to neglect excitonic effects, and to examine our theoretical findings and make the analysis as general as possible we here cross-check the results with data obtained from the dielectric permittivity of CrI$_3$ based on recently reported $GW$-BSE study \cite{wu2019physical}. The relationship between permittivity and optical conductivity is well known and straight forward hence values of the conductivity needed for solving a set of Maxwell's equations can be extracted from $GW$-BSE calculations. 

We simulate the propagation of a TM-polarized electromagnetic field at $\omega=1.4$~eV (within $GW$-BSE approach this energy is slightly below the first bright exciton level positioned at $\omega_A=1.5$~eV and the effects due to excitons should not play a significant role, making a direct comparison with GGA$+U$ possible) in a nanolayer of CrI$_3$ of thickness $d=300$~nm in the Kretschmann-Raether (KR) configuration. The results of numerical simulations featuring the distribution of $y$-component of the magnetic field, $H_y$, in the {\it{xz}}-plane clearly demonstrate the propagation of surface electromagnetic waves bound by a CrI$_3$ nanolayer. In the reflectance spectrum, plotted in Fig.~\ref{fig:reflectance}, the dip corresponding to the excitation of the surface plasmon polaritons appears at the incident angle $\alpha=50.4^{\circ}$, which qualitatively agrees with the results obtained based on GGA$+U$ study, albeit with a somewhat higher value compared to the angle of total internal reflection discussed above. The latter unambiguously suggests that peculiarities associated with surface plasmon polariton generation are robust with respect to the computational scheme in use.

\subsection{The symmetric bilayer structure} 
\begin{figure}[h!]
\centering
  \includegraphics[scale=0.5]{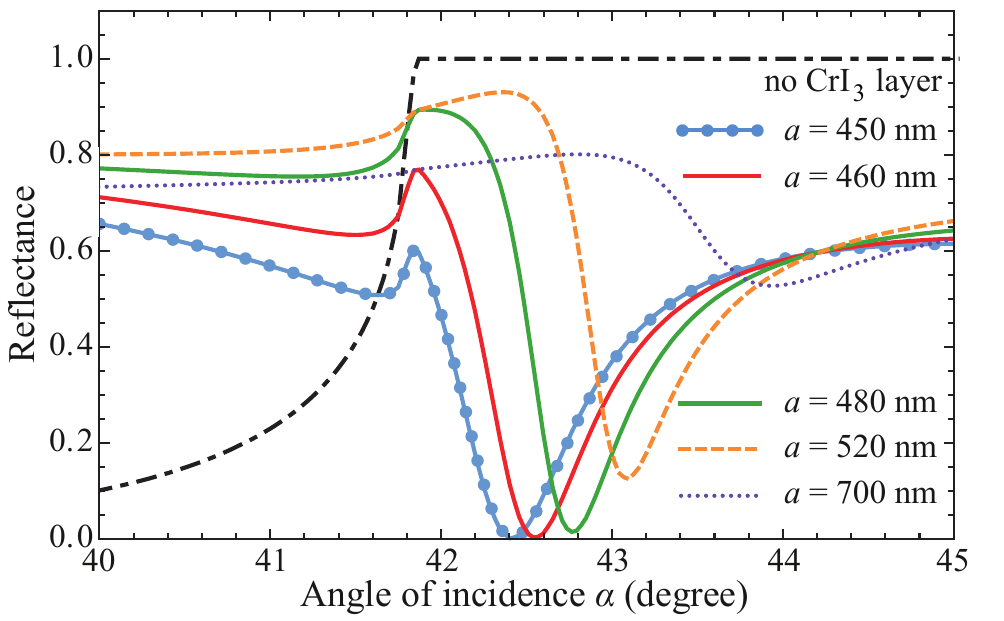}
  \caption{Angular response of the system composed by two CrI$_3$ layers of thickness $d=100$ nm separated by SiO$_2$ layer with thickness $a$ and refractive index $n_{\mathrm{SiO}_2}=1.44$ (Fig.~\ref{fig:model}b). The light enters the system via the glass prism at the angle $\alpha$. By varying the spacer thickness, the position of the reflectance minimum shifts from $42.4^{\circ}$ at $a=450$ nm (blue solid-circled line) to $43.1^{\circ}$ at $a=520$ nm (orange dashed line). All curves show a characteristic feature of total internal reflection as a local maximum at $\alpha=41.8^{\circ}$. When the guided mode condition on the layer thickness is not fulfilled (e.g. when $a=700$ nm) the surface plasmon resonance vanishes from the spectrum (violet dotted line)}.
  \label{fig:thickness}
\end{figure}

We also investigate the performance of a surface plasmon resonance in a symmetric bilayer structure, where a SiO$_2$ layer with thickness $a$ is sandwiched between two identical layers CrI$_3$ of thickness $d$=100 nm (shown in Fig.~\ref{fig:model}b) within GGA+$U$ approach. We find that when light enters the system under the incident angle $\alpha$ the surface plasmons can be created on the interface between CrI$_3$ and air, and the position of the resonance angle increases in the range from $42.4^{\circ}$ to $43.1^{\circ}$ for a variation of spacer thickness (SiO$_2$ layer)  between 450 and  520 nm. The results of the calculations are shown in Fig.~\ref{fig:thickness}. Note that for the excitation of surface plasmons at the boundary between CrI$_3$ layer and air, the electromagnetic wave should propagate through the spacer layer in the form of a guided mode that at the same time leads to the generation of an evanescent wave in the neighbouring low-refractive index medium. Therefore, it is clear that for plasmon generation, the spacer thickness $a$ is optimal when it meets the value of any supported wave guide mode in the SiO$_2$ spacer layer. Meanwhile, when this condition is not fulfilled (for e.g. $a=700$ nm, see violet dotted line in Fig.~\ref{fig:thickness}), the surface plasmon resonance vanishes from the spectrum.

\section{Conclusions} 
In this paper, we addressed the properties of surface electromagnetic waves bound by a thin  ferromagnet layer of chromium triiodide in the Kretschmann-Raether configuration. Using the conductivity tensor obtained within the Kubo formalism from the {\it ab initio} calculations, we evaluated the dispersion relation of surface electromagnetic wave mode in CrI$_3$ monolayer. By a direct numerical solution to a set of Maxwell's equations we showed that in a rather large energy window these waves can be stabilized. Considering thin chromium triiodide layer we have estimated the critical angle which corresponds to the absorption level of charge density waves in CrI$_3$. It turns out that this state inherits the properties of TM-surface plasmon polariton and its features are controlled by the thickness of CrI$_3$ layer, leading to the decrease in the effect with increasing layer thickness. We also examined the process of surface plasmon generation in the bilayer CrI$_3$ structures by considering the CrI$_3$/SiO$_2$/CrI$_3$ system when the light is injected to the system by a glass prism. We have shown that the angular position of the surface plasmon dip on the reflectance spectrum and its width strongly depend on the thickness of the SiO$_2$ layer. We concluded that for surface plasmon generation the thickness of the spacer layer should support one of the guided modes in the system under consideration. When the imposed condition is not fulfilled, the surface plasmons are not excited at the interface between CrI$_3$ and air. We believe that the results of this theoretical analysis will trigger further experimental activity with new materials in the field of nanomagnetoplasmonics.

\section*{Funding}

Russian Foundation for Basic Research (RFBR) (19-32-60020, 20-52-S52001); Swedish Research Council (Vetenskapsr{\aa}det)  (2018-04383, 2015-04608, 2016-05980, 2019-05304); Ministry of Science and Technology and the National Center for Theoretical Sciences, Taiwan; Knut and Alice Wallenberg foundation (KAW) 2018.0060; SERC; eSSENCE; Academia Sinica Thematic Research Program (AS-106-TP-A07).

\section*{Disclosures}
The authors declare no conflicts of interest.

\bibliography{manuscript.bbl}

\end{document}